\documentclass[12pt]{article}
\usepackage{graphicx}
\usepackage{type1cm, amsmath, amssymb}
\makeatletter
\newlength\captionwidth \captionwidth=\hsize

\def\hangcaption{\refstepcounter\@captype\@dblarg{\@hangcaption\@captype}}

\long\def\@hangcaption#1[#2]#3{%
  \baselineskip 14.5\p@
  \addcontentsline{\csname ext@#1\endcsname}{#1}%
    {\protect\numberline{\csname the#1\endcsname}{\ignorespaces #2}}%
  \par
  \begingroup
    \@parboxrestore
    \setbox\@tempboxa\hbox{\csname fnum@#1\endcsname: #3}%
    \ifdim \wd\@tempboxa > \captionwidth
      \begingroup
        \setbox\@tempboxa\hbox{\csname fnum@#1\endcsname: }%
        \advance \captionwidth by -\wd\@tempboxa
        \@makecaption{\csname fnum@#1\endcsname}%
            {\ignorespaces{\protect\parbox[t]{\captionwidth}{\leavevmode#3}}}%
      \endgroup
    \else
      \begingroup
        \@makecaption{\csname fnum@#1\endcsname}{\ignorespaces #3}%
      \endgroup
    \fi
    \par\vspace{1.7ex}%
  \endgroup}

\let\isucaption\hangcaption
\makeatother

\renewcommand{\caption}{\isucaption}

\newcommand{\nn}{\nonumber}

\newcommand{\Rb}{\mathbb{R}}
\newcommand{\Cb}{\mathbb{C}}
\newcommand{\Ncal}{\mathcal{N}}
\newcommand{\del}{\partial}

\newcommand{\vp}{\varphi}
\DeclareMathOperator*{\Tr}{{\rm Tr}}
\DeclareMathOperator*{\tr}{{\rm tr}}

\newcommand{\rap}[2]
{\setbox1=\hbox{#1}%
\setbox2=\hbox to\wd1{\hss #2\hss}%
\mbox{\rlap{\box1}\box2}}
\newcommand{\sla}[1]{\rap{$#1$}{/}}

\newcommand{\delsla}{\sla{\del}}

\newcommand{\Fh}{\widehat{F}}
\newcommand{\Kc}{\check{K}}

\newcommand{\Jt}{\widetilde{J}}
\newcommand{\Lt}{\widetilde{L}}
\newcommand{\Kt}{\widetilde{K}}
\newcommand{\Ft}{\widetilde{F}}
\newcommand{\Fs}{\sla{F}}
\newcommand{\Ks}{\sla{K}}
\newcommand{\Ls}{\sla{L}}
\newcommand{\Js}{\sla{J}}
\newcommand{\dt}{\tilde{d}}
\newcommand{\delt}{\tilde{\del}}

\newcommand{\sigmac}{\check{\sigma}}
\newcommand{\sigc}{\sigma_{C}}
\newcommand{\sigmah}{\hat{\sigma}}
\newcommand{\sigmat}{\tilde{\sigma}}

\newcommand{\chic}{\check{\chi}}
\newcommand{\chih}{\hat{\chi}}
\newcommand{\nablac}{\overset{\circ}{\nabla}}

\numberwithin{equation}{section}

\begin{document}

%%%%%%%%%%%%%%%%%%%%%%%%%%%%%% Title page %%%%%%%%%%%%%%%%%%%%%%%%%%%%%%%%%%%%%
\thispagestyle{empty}

\begin{flushright}
 \begin{tabular}{l}
 {\tt hep-th/0601089}\\
 IHES/P/06/01
 \end{tabular}
\end{flushright}

 \vfill
 \begin{center}
{\LARGE
 \centerline{Bubbling Geometries for Half BPS Wilson Lines} 
 \vskip 3mm
 \centerline{ }
% \vskip 2.5 truecm
}

 \vskip 2.0 truecm
\noindent{ \large  Satoshi Yamaguchi} \\
{\sf yamaguch@ihes.fr}
\bigskip

 \vskip .6 truecm
 {
 {\it 
IHES, Le Bois-Marie, 35, route de Chartres\\ 
F-91440 Bures-sur-Yvette,
FRANCE
} 
 }
 \vskip .4 truecm

 \end{center}

 \vfill
\vskip 0.5 truecm

\begin{abstract}
We consider the supergravity backgrounds that correspond to supersymmetric Wilson line operators in the context of AdS/CFT correspondence. We study the gravitino and dilatino conditions of the IIB supergravity under the appropriate ansatz, and obtain some necessary conditions for a supergravity background that preserves the same symmetry as the supersymmetric Wilson lines. The supergravity solutions are characterized by continuous version of maya diagrams. This diagram is related to the eigenvalue distribution of the Gaussian matrix model. We also consider the similar backgrounds of the 11-dimensional supergravity.
\end{abstract}
\vfill
\vskip 0.5 truecm

\newpage
\setcounter{page}{1}
%\tableofcontents

%%%%%%%%%%%%%%%%%%%%%%%%%%%%% end of title page %%%%%%%%%%%%%%%%%%%%%%%%%%%%%%%
\section{Introduction}

In the gauge theory, the most important class of non-local operators is the Wilson loop operators. Especially, in $\Ncal=4$ super Yang-Mills theory, the Wilson loops with the scalar term are the most fundamental ones. The form of this class of operators is
\begin{align}
 W_{R}(C)=\Tr_{R}\left[P\exp\left(i\oint_{C} ds (A_{\mu}\dot{x}^{\mu}+\vp_4 
 |\dot{x}|)\right)\right], \label{wilson-loop}
\end{align}
where $\dot{x}^{\mu}$ denotes $\del x^{\mu}/\del s$, and $\vp_4$ is one of the six real scalar fields in the $\Ncal=4$ super Yang-Mills theory. The label $R$ denotes the representation of the gauge group, in which the trace is taken
\footnote{$\Tr_R$ denotes a trace in the representation $R$, while $\tr$ denotes a trace in the fundamental representation.}.

In the context of AdS/CFT correspondence, the Wilson loops are described by the fundamental strings in $AdS$ spacetime \cite{Rey:1998ik,Maldacena:1998im}. The vacuum expectation value of the Wilson loop has been calculated as the on-shell action of the fundamental string. In \cite{Drukker:2005kx}, the circular Wilson loop has been calculated as an on-shell action of D3-brane. This calculation by D3-brane includes some higher genus corrections.

If the gauge group is $SU(N)$, the representation label $R$ in \eqref{wilson-loop} is expressed by a Young diagram. In the probe picture---fundamental strings or D3-branes---, we can only see the special kind of representations. How can we see the Wilson loops with arbitrary Young diagrams in the AdS side?

Our approach in this paper to this problem is the similar one to the work of Lin, Lunin, Maldacena \cite{Lin:2004nb}. They have constructed supergravity solutions which correspond to half BPS local operators. In their solution, one can see the phase space of free fermions in the harmonic potential. This picture is consistent with the dynamics of eigenvalues of the Gaussian matrix quantum mechanics proposed by \cite{Corley:2001zk,Berenstein:2004kk,Takayama:2005yq}.

In this paper, we study the supergravity solutions corresponding to the half BPS straight Wilson lines. We consider the supersymmetry conditions under the appropriate ansatz, and derive some necessary conditions.

\begin{figure}[bpht]
\begin{center}
   \includegraphics[width=300pt]{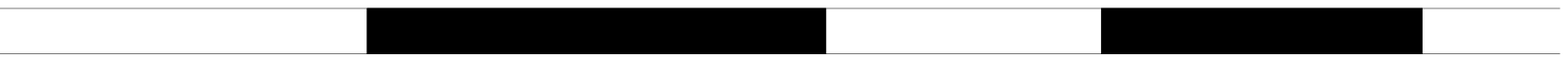}	       
\end{center}
\caption{The black and white pattern that appears on the boundary of the base 2-dimensional space. At a black point, $S^2$ shrinks, and at the white point $S^4$ shrinks.}
\label{fig-maya}
\end{figure}

\begin{figure}
\begin{center}
   \includegraphics[width=300pt]{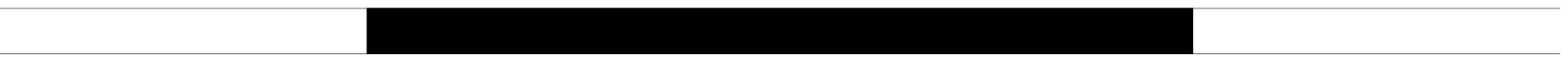}	       
\end{center}
\caption{The pattern that characterizes the $AdS_5\times S^5$ solution. The black segment expresses $S^5$.}\label{fig-ads}
\end{figure}

Let us summarize the result of this paper here. From the necessary conditions, we find the smooth supergravity solution is characterized by 1-dimensional black and white pattern like figure \ref{fig-maya}.
For example, $AdS_5\times S^5$ is characterized by the pattern of figure \ref{fig-ads}. 
If this kind of pattern is discretized in some mechanism, we get a ``maya diagram.'' The maya diagram is known to be an equivalent label as a Young diagram. Actually, due to the quantization of the flux, the black and white pattern is discretized. When we take the correspondence to the Gaussian matrix model into account, we will find that the Young diagram obtained from this maya diagram corresponds to the Wilson line of the monomial basis, not the representation basis.

The correspondence to the matrix model can be seen as follows. It is conjectured that the expectation value of a circular Wilson loop is calculated by the Gaussian matrix model \cite{Erickson:2000af,Drukker:2000rr}. {\em We claim that the black and white pattern in the geometry corresponds to the eigenvalue distribution of the matrix model.} The black part is where the eigenvalues exist.

We checked that the length of the black segment in the $AdS_5\times S^5$ is quantized like $\sqrt{N}$, which is just the same as the length of the cut of the Gaussian matrix model. As a father check, we compare an $AdS_2\times S^2$ D3-brane probe of \cite{Drukker:2005kx} and the eigenvalue distribution of the saddle point of the integral $\left\langle \frac{1}{N}\tr[e^{kM}]\right\rangle_{\text{mm}}$ in the matrix model. It is shown that they are completely consistent. This identification leads to ``D-brane exclusion principle''; Two same $AdS_2\times S^2$ D3-branes cannot exist at the same position at the same time.

We also investigate a similar problem in M-theory. There are half BPS surface operators in the 6-dimensional (2,0) SCFT. Also in the 3-dimensional $\Ncal=8$ SCFT there are half BPS surface operators. We derive some necessary conditions for the supergravity solutions that correspond to these surface operators.

The construction of this paper is as follows. In section \ref{IIB}, we study the supersymmetry conditions in IIB supergravity under the ansatz, and derive some necessary conditions. In section \ref{interpretation}, we see the black and white pattern on the boundary of the 2-dimensional base space. We compare it to the eigenvalue distribution of the Gaussian matrix model. In section \ref{M} we investigate the similar problem in M-theory. Section \ref{conclusion} is devoted to conclusions and discussions.

\section{1/2 BPS Wilson lines and Bubbling Geometry}
\label{IIB}
In this section, we investigate the IIB geometry that corresponds to half BPS straight Wilson lines. First we consider the symmetry preserved by the Wilson lines and make the ansatz. Next, we study the supersymmetry conditions and derive some necessary conditions.

\subsection{Symmetry and Ansatz for Wilson lines}

In this subsection, we consider the symmetry preserved by the Wilson line operators and make the ansatz respecting that symmetry. We are considering the straight Wilson line which extends to the time direction. It is equivalent to introducing a test particle with infinite mass, sitting on the origin of the space. For this kind of Wilson line, \eqref{wilson-loop} can be written as
\begin{align}
 W_{R}=\Tr_{R}\left[P\exp\left(i\int dx^{0} (A_0+\vp_4 
 )\right)\right]. \label{line0}
\end{align}
Let us see the bosonic symmetry which leaves this operator invariant.

First, the Wilson line looks like a particle sit on the origin of the space. Therefore, it preserves the rotation around the origin. This group is SO(3).

Next, this Wilson line is invariant under the time translation. This Wilson line is also scale invariant, namely, it preserves the dilatation symmetry as well as the special conformal symmetry of the time direction. These three generators of the transformations make the algebra SL(2,R).

Finally, the Wilson line operator \eqref{line0} preserves SO(5) part of the R-symmetry, because this operator includes only $\vp_4$ but does not include the other five scalar fields $\vp_5,\dots,\vp_9$.

In summary, the total bosonic symmetry is SL(2,R)$\times$SO(3)$\times$SO(5). The spacetime part of this symmetry SL(2,R)$\times$SO(3) can be seen more clearly by Weyl transformation\cite{Kapustin:2005py}. The 4-dimensional Minkowski space can be transformed by Weyl transformation to $AdS_2\times S^2$. The isometry of $AdS_2\times S^2$ is SL(2,R)$\times$SO(3).

Let us turn to making the ansatz that respects the symmetry SL(2,R)$\times$SO(3)$\times$SO(5). First, the metric can be written in the form
\begin{align}
 ds^2=e^{2A}d\check{\Omega}_2^2+ds_2^2+e^{2B}d \hat{\Omega}_2^2
  +e^{2C}d\Omega_4^2, \label{metric-ansatz}
\end{align}
where $d\check{\Omega}_2^2,d \hat{\Omega}_2^2,d\Omega_4^2$ are the metrics of unit $AdS_2$, $S^2$, $S^4$ respectively. $ds_2^2$ in \eqref{metric-ansatz} is a general 2-dimensional metric to be determined. Let us call this 2-dimensional space expressed by $ds_2^2$ ``base space.'' $A,B,$ and $C$ in \eqref{metric-ansatz} are functions on the base space.

Next, let us consider the ansatz for the form fields. We first write here some notations. Let $(\theta^{0},\theta^{1})$ be the vielbein of unit $AdS_2$, $(\theta^4,\theta^5)$ be the vielbein of unit $S^2$, and $(\theta^6,\dots,\theta^9)$ be the vielbein of unit $S^4$. The vielbein of the metric \eqref{metric-ansatz} are denoted by $E^{M},\ M=0,1,\dots,9$. This means $E^0=e^A \theta^{0},\ E^1=e^A \theta^{1}$, and $E^4=e^B \theta^{4},\ E^5=e^B \theta^{5}$, and $E^6=e^C \theta^{6},\dots , E^9=e^C \theta^{9}$.

The most general ansatz for the form fields that respects the SL(2,R)$\times$SO(3)$\times$SO(5) symmetry is as follows.
\begin{align}
 &H_3=F E^0 E^1+\Fh E^4 E^5,\\
 &G_3=4e^{-\phi}(\Kc E^0 E^1+ K E^4 E^5),\\
 &G_5=4e^{-\phi}(L E^0 E^1 E^4 E^5+\Lt E^6 E^7 E^8 E^9),
\end{align}
where $F,\Fh,\Kc,K,L,\Lt$ are 1-forms on the base 2-dimensional space. The RR 1-form $G_1$ should also be a 1-form on the base space. The dilaton should be a function on the base space. The self-duality of $G_5$ implies $L$ and $\Lt$ are
dual to each other, namely,
\begin{align}
 \Lt_2=-L_3,\qquad L_2=\Lt_3.
\end{align}
We use the tilde symbol $\widetilde{\ \ }$ as 2-dimensional Hodge dual in the rest of this paper.

If all of these fields are excited, the background may include not only Wilson lines but also 'tHooft lines. In this paper, we concentrate on the Wilson lines. This means that the test particle has only electric charges but does not have magnetic charges. Hence we have to consider how to truncate the fields of supergravity. First of all, since the original background $AdS_5\times S^5$ contains $L$ and $\tilde L$ excitations, these two fields cannot be set to be 0. Next, in the probe picture, the Wilson line is a fundamental string or a D3-brane with electric flux. In both cases, the brane has the charges for the NSNS 3-form field strength along the $AdS_2$. This means $F$ should not be 0. $\Lt$ and $F$ excitations may become a source for the field $K$. Actually, the equation of motion for $K$ is written as
\begin{align}
 e^{-2B-4C+\phi}d(e^{2B+4C-\phi} \Kt)=-\Lt F.
\end{align}
As a result, we can put $0=G_1=\Fh=\Kc=0$ consistently. 
The unknown fields on the 2-dimensional base space are the following fields.
\begin{itemize}
 \item Metric $ds_2^2$.
 \item Scalar fields $A,B,C,\phi$.
 \item 1-forms $F,K,L$.
\end{itemize}
These fields should be determined by the supersymmetry conditions and the equations of motion. 

\subsection{Analysis of Supersymmetry conditions}
In this subsection, we consider the supersymmetry conditions under the ansatz that we put in the previous subsection. These conditions lead to the necessary conditions for the backgrounds.

Let us first prepare the things needed to write down the supersymmetry conditions in terms of the base 2-dimensional space. We use the set of the 10-dimensional gamma matrices.
\begin{align}
 &\Gamma^0=\sigmac^{0}\otimes \sigc \otimes 1 \otimes 1,
 &&\Gamma^1=\sigmac^{1}\otimes \sigc \otimes 1 \otimes 1,\nn\\
 &\Gamma^2=1\otimes \sigma_1\otimes 1 \otimes 1,
 &&\Gamma^3=1\otimes \sigma_2\otimes 1 \otimes 1,\nn\\
 &\Gamma^4=\sigmac^{3}\otimes \sigc \otimes \sigmah_4 \otimes 1,
 &&\Gamma^5=\sigmac^{3}\otimes \sigc \otimes \sigmah_5 \otimes 1,\nn\\
 &\Gamma^a=\sigmac^{3}\otimes \sigc \otimes \sigmah_6 \otimes \gamma^{a},
&&\qquad (a=6,7,8,9),
\end{align}
where $(\sigma_1,\sigma_2,\sigc)$ and $(\sigmah_4,\sigmah_5,\sigmah_6)$ are sets of Pauli matrices. $(\sigmac_1,\sigmac_2,\sigmac_3)$ is another set of Pauli matrices, and we define $\sigmac^{0}$ as $\sigmac^{0}:=i\sigmac_2$. $\gamma^{a}, \ (a=6,7,8,9)$ are gamma matrices of Euclidean 4-dimensions.

We also use some typical spinors in AdS spacetimes and spheres, which is called ``Killing spinors.'' For example in $AdS_2$, we have spinors satisfying the relation
\begin{align}
 &\nablac_p \chic^{I}_{a}=\frac i2 a \sigmac_{p}\chic^{I}_{-a},\qquad 
 \sigmac_{3}\chic^{I}_{a}=a\chic^{I}_{a},\qquad
  (p=0,1,\qquad a=\pm 1,\qquad I=1,2),
\end{align}
where $\nablac$ is the covariant derivative of Levi-Civit\'a connection of unit $AdS_2$. As the same way, there are Killing spinors in $S^2$ and $S^4$.
\begin{align}
  &\nablac_p \chih^{J}_{b}=\frac 12 b \sigmah_{p}\chih^{J}_{-b},\qquad 
 \sigmah_{6}\chih^{J}_{b}=b\chih^{J}_{b},\qquad
 (p=4,5,\qquad b=\pm 1,\qquad J=1,2),\\
  &\nablac_p \chi^{K}_{c}=\frac 12 c \gamma_{p}\chi^{K}_{-c},\qquad 
 \gamma_{6789}\chi^{K}_{c}=c\chi^{K}_{c},\qquad
 (p=6,\dots,9,\qquad c=\pm 1,\qquad K=1,2,3,4).
\end{align}

One can reduce the problem to 2-dimensions by expanding the 10-dimensional spinor pair $\xi$ by the above Killing spinors.
\begin{align}
 \xi=\sum_{a,b,c}\chic^{I}_{a}\otimes\epsilon_{abcIJK}\otimes\chih^{J}_{b}
\otimes \chi^{K}_{c}.
\end{align}
Each $\epsilon_{abcIJK}$ is a pair of 2-dimensional spinors. $\Gamma_2,\Gamma_3,\sigc,\tau_1,\tau_2,\tau_3$ act on $\epsilon_{abcIJK}$.

By using these materials, we can reduce the problem to the base 2-dimensional space. The supersymmetry conditions\eqref{gravitino-condition},\eqref{dilatino-condition} can be written in terms of 2-dimensional language as
\begin{align}
 e^{-A}\epsilon_{(-a)bc}&=\left[
-ibc\delsla A -\frac i2 abc \Fs \tau_3 +iac\Ls \tau_2 +c\Ks \tau_1
\right]\epsilon_{abc},\nn\\
 e^{-B}\epsilon_{a(-b)c}&=\left[
c\delsla B -abc\Ls \tau_2 -ibc\Ks \tau_1
\right]\epsilon_{abc},\nn\\
 e^{-C}\epsilon_{ab(-c)}&=\left[
\delsla C +ab\Ls \tau_2 +ib\Ks \tau_1
\right]\epsilon_{abc},\nn\\
\nabla_{m}\epsilon_{abc}&=
\left[
-\frac14 a F_m \tau_3+\frac12(ab \Ls \tau_2+ib \Ks \tau_1)\Gamma_m
\right]\epsilon_{abc},\nn\\
0&=\left[
\delsla \phi+\frac12 a \Fs \tau_3 + 2ib\Ks \tau_1
\right]\epsilon_{abc}.\label{spi0}
\end{align}
In these equations, we neglect the $I,J,K$ indices.

It is convenient to introduce some other sets of Pauli matrices to express \eqref{spi0} in the simple way. Let $\mu_j,\nu_j,\lambda_j,\ (j=1,2,3)$ are sets of Pauli matrices which act on indices $a,b,c$ respectively. This means, for example,
\begin{align}
 (\mu_j \epsilon)_{abc}=(\mu_j)_{aa'}\epsilon_{a'bc}.
\end{align}
We also define a set of matrices $\rho_j,\ (j=1,2,3)$ as
\begin{align}
 \rho_1=\mu_3\tau_3,\qquad \rho_2=\nu_3\tau_1,\qquad \rho_3=\mu_3\nu_3\tau_2.
\end{align}
These three matrices $\rho_j$ satisfy the algebra of a set of Pauli matrices.

Using these notations, we can express \eqref{spi0} in more convenient way.
\begin{align}
 -ie^{-A}\mu_1\nu_3\lambda_3\epsilon
&=\left[-\delsla A -\frac12 \rho_1 \Fs +\rho_3 \Ls-i\rho_2 \Ks\right]\epsilon,
\label{a}\\
e^{-B}\nu_1\lambda_3\epsilon&=\left[
\delsla B-\rho_3\Ls -i\rho_2 \Ks\right]\epsilon,\label{b}\\
e^{-C}\lambda_1\epsilon&=\left[
\delsla C + \rho_3\Ls +i\rho_2\Ks\right]\epsilon,\label{c}\\
\nabla_m\epsilon&=\left[
-\frac14\rho_1F_m+\frac12 \rho_3 \Ls \Gamma_m+\frac i2 \rho_2 \Ks\Gamma_m
\right]\epsilon,\label{m}\\
0&=\left[
\delsla \phi+\frac12\rho_1\Fs + 2i\rho_2\Ks\right]\epsilon.\label{phi}
\end{align}
The Weyl condition $\Gamma_{10}\xi=\xi$ can be written in terms of $\epsilon$
as
\begin{align}
 \Gamma_{23}\epsilon = \eta \epsilon,\qquad \eta:=-i\mu_3\nu_3\lambda_3.
\end{align}

The standard method to solve the supersymmetry conditions is making the spinor bilinears and considering the differential equation for those bilinears. Here is a part of the relevant bilinears.
\begin{align}
 f_0=\epsilon^{\dag}\epsilon,\qquad f_j=\epsilon^{\dag}\rho_j\epsilon,\nn\\
 g_0=i\epsilon^{\dag}\eta\epsilon,\qquad g_j=i\epsilon^{\dag}\eta\rho_j\epsilon.
\label{bilinears}
\end{align}
$f_j$ and $g_j$ are real functions on the base 2-dimensional space.

The derivative of these bilinears can be calculated by using \eqref{m} as
\begin{align}
 df_0&=-\frac12 f_1 F+f_3 L+g_2 \Kt, \label{df0}\\
 df_1&=-\frac12 f_0 F-g_2 \Lt- f_3 K,\\
 dg_2&=f_1 \Lt+f_0 \Kt,\label{dg2}\\
 df_3&=f_0 L+f_1 K.\label{df3}
\end{align}
Here, we use tilde notation as the 2-dimensional Hodge dual. For example,
\begin{align}
 \Kt_m=-\varepsilon_{mn}K_n,\qquad \varepsilon_{23}=+1,\qquad m,n=2,3.
\end{align}

One can also derive some relations among the bilinears from \eqref{a}-\eqref{c}. For example, \eqref{a} multiplied by $\epsilon^{\dag}\Gamma_m$ reads
\begin{align}
 \epsilon^{\dag}\Gamma_{m}(-i)e^{-A}\mu_1\nu_3\lambda_3\epsilon
&=\epsilon^{\dag}\Gamma_{m}\left[-\delsla A -\frac12 \rho_1 \Fs +\rho_3 \Ls-i\rho_2 \Ks\right]\epsilon,
\label{a1}
\end{align}
and the hermitian conjugation of this equation reads
\begin{align}
 \epsilon^{\dag}(+i)e^{-A}\mu_1\nu_3\lambda_3\Gamma_{m}\epsilon
&=\epsilon^{\dag}\left[-\delsla A -\frac12 \rho_1 \Fs +\rho_3 \Ls+i\rho_2 \Ks\right]\Gamma_{m}\epsilon,
\label{a2} 
\end{align}
If we add eq.\eqref{a1} and eq.\eqref{a2}, we obtain the relation
\begin{align}
 f_0dA&=-\frac12 f_1 F+f_3 L+g_2 \Kt
\end{align}
Comparing this equation and eq.\eqref{df0}, we can conclude that $e^{A}$ is proportional to $f_0$. We have the freedom of the normalization of $\epsilon$, so we will fix this normalization by $f_0=e^{A}$.

As the same way, one can derive, from \eqref{b} and \eqref{c}, the relations
\begin{align}
 f_3dB=f_0 L+f_1 K,\qquad
 g_2dC=f_1 \Lt+f_0 \Kt,
\end{align}
and comparing these equations to \eqref{df3} and \eqref{dg2}, we can conclude that $e^{B}$ is proportional to $f_3$ and $e^{C}$ is proportional to $g_2$.

Can we determine the coefficient of the proportions? Actually, in the $AdS_5\times S^5$ solution, $e^{B}=f_3$ and $e^{C}=g_2$ are satisfied. The solution we want is asymptotically $AdS_5\times S^5$. So $e^{B}=f_3$ and $e^{C}=g_2$ should be satisfied in our solution.

To proceed the analysis, let us consider the spinor bilinear $S_m$ defined as
\begin{align}
 S_m:=\epsilon^{\dag}\Gamma_{m}(-i\rho_2)\nu_1\lambda_3\epsilon
 =\epsilon^{\dag}\Gamma_{m}\nu_2\lambda_3\tau_1\epsilon.
\end{align}
The derivative of $S_m$ can be calculated by using \eqref{m}.
\begin{align}
 \nabla_{n}S_{m}=\epsilon^{\dag}\nu_2\lambda_3\tau_1\left[
\frac12 \rho_3 \Gamma_{m}\Ls \Gamma_{n}+\frac{i}{2}\rho_2 \Gamma_m\Ks \Gamma_{n}\right]
\epsilon+(m\leftrightarrow n).
\end{align}
Especially, $\nabla_n S_{m}$ is symmetric under the exchange of $m$ and $n$. That means, 1-form $S:=S_m E^{m}$ is closed.

On the other hand, $S_m$ is related to other bilinears as follows. When we multiply $\epsilon^{\dag}\Gamma_{m}(-i\rho_2)$ to \eqref{b}, we obtain the relation
\begin{align}
 e^{-B}S_{m}=-g_2\delt_m B + f_1 L_m+f_0 K_m.
\end{align}
We can simplify the above relation by Hodge dual of \eqref{dg2} and $g_2=e^{C}$ as
\begin{align}
 S=-\dt e^{B+C}.
\end{align}
As we saw before, $S$ is a closed 1-form. So we can conclude that $d \dt e^{B+C}=0$. In other words, $e^{B+C}$ is a harmonic function on the base 2-dimensional space.

Now, we can take the coordinates of the base 2-dimensional space. One coordinate is $y=e^{B+C}$. As we have shown, $\dt y$ is a closed one form. Thus we can define at least locally the other coordinate $x$ orthogonal to $y$, namely $dx:= \dt y$.

To determine the metric of the base 2-dimensional space, let us consider the norm of $dy$. If we add \eqref{b} and \eqref{c}, we will obtain
\begin{align}
 \left[e^{-B}\nu_1\lambda_3+e^{-C}\lambda_1-\delsla(B+C)\right]\epsilon=0.
\end{align}
Multiplying this equation $\left[e^{-B}\nu_1\lambda_3+e^{-C}\lambda_1+\delsla(B+C)\right]$ reads $|d(B+C)|^2=e^{-2B}+e^{-2C}$. Consequently the norm of $dy=de^{B+C}$ can be written as $|dy|^2=e^{2B}+e^{2C}$. Now, we can express the metric of the base 2-dimensional space in the simple form
\begin{align}
 ds_2^2=\frac{1}{e^{2B}+e^{2C}}(dy^2+dx^2).\label{2metric}
\end{align}

From the dilatino condition \eqref{phi}, by multiplying $\epsilon^{\dag}\Gamma_m$ and $\epsilon^{\dag}\Gamma_m\rho_{j}$, we obtain the following set of relations
\begin{align}
 &0=f_0 d\phi+\frac12 f_1 F - 2g_2 \Kt,\qquad
  0=f_1 d\phi+\frac12 f_0 F - 2f_3 K,\nn\\
 &0=g_2 d\phi+\frac12 f_3 \Ft-2f_0 \Kt,\qquad
  0=f_3 d\phi-\frac12 g_2\Ft +2f_1 K.\label{di}
\end{align}
These relations lead to the algebraic relation among $f_0,f_1,g_2,f_3$ as
\begin{align}
 -f_0^2+f_1^2+g_2^2+f_3^2=0.
\end{align}

In summary, we can express every unknown quantity in terms of $A,B,C$, by using \eqref{df0}-\eqref{df3},\eqref{2metric}, and \eqref{di}
\begin{align}
 &ds_2^2=\frac{1}{e^{2B}+e^{2C}}(dy^2+dx^2),\label{sol1}\\
 &F=\frac{2}{f_{0}^2-f_1^2}\left[
  f_1 df_0-f_0df_1
 +f_3\dt g_2-g_2 \dt f_3\right],\\
 &K=\frac{1}{f_{0}^2-f_1^2}\left[
   -f_0 \dt g_2-f_1 d f_3\right],\\
 &L=\frac{1}{f_{0}^2-f_1^2}\left[
   f_1 \dt g_2+f_0 d f_3\right],\\
 &d\phi=\frac{2}{(f_{0}^2-f_1^2)^2}\left[f_0^2 g_2d g_2+f_1^2 f_3 d f_3
   +f_0f_1(f_3 \dt g_2 -g_2 \dt f_3)\right],\label{sol2}
\end{align}
where 
\begin{align}
 y=e^{B+C},\qquad f_0=e^{A},\qquad f_3=e^{B}, \qquad g_2=e^{C}, 
\qquad f_1=\sqrt{e^{2A}-e^{2B}-e^{2C}}.
\end{align}

This condition is necessary but not sufficient. For example $dd\phi=0$ and \eqref{sol2} implies a differential equation for $A,B,C$. The Bianchi identities and equation of motions for the form fields are written as
\begin{align}
 2 dA F+ dF=0,\qquad (2dB-d\phi)K+dK=0,\nn\\
 (2dA+2dB-d\phi)L+dL= F K,\qquad
 (4dC-d\phi)\Lt+d\Lt=0,\nn\\
 (2dB+4dC-2d\phi)\Ft+d\Ft=16K\Lt,\qquad
 (2dA+4dC-d\phi)\Kt+d\Kt=-\Lt F. 
\end{align}
These equations also lead to some differential equations for $A,B,C$.
It is a future problem to obtain the independent set of differential equations and show it is sufficient.

\section{Interpretation of the geometry and the Gaussian matrix model}
\label{interpretation}
In this section, we study the detail of the geometry which we obtain in the previous section. We will find the continuous version of maya diagram at the boundary of the base 2-dimensional space. We also compare this pattern with the eigenvalue distribution of the Gaussian matrix model.

\subsection{The structure of the geometry}
Let us see the detail of the geometry \eqref{sol1}-\eqref{sol2}.

As the same way as in \cite{Lin:2004nb}, one of the coordinate $y$ is the product of the radii of $S^2$ and $S^4$. So $y$ is greater than $0$ and the line $y=0$ is the boundary of the base 2-dimensional space. On this boundary(let us call it $x$-axis),  $e^{B}=0$ or $e^{C}=0$ should be satisfied because $y:=e^{B}e^{C}=0$. Actually at a point on the $x$-axis, if either $e^{B}$ or $e^{C}$ vanish and the other remain finite, the geometry is smooth at the point. For example, near the point where $e^{B}=0$ and $e^{C}$ is finite, the relevant part of the metric can be written as
\begin{align}
 ds^2=e^{-2C}dy^2+e^{-2C}y^2 d\hat{\Omega}_2^2+\dots.
\end{align}
This metric is smooth at $y=0$.

If the geometry is given, we can draw a 1-dimensional black and white pattern like figure \ref{fig-maya} as follows. Take a point on the $x$-axis. If at that point $e^{B}=0$ and $e^{C}\ne 0$ are satisfied, mark that point by black. If $e^{C}=0$ and $e^{B}\ne 0$, mark that point by white. Then one obtains the 1-dimensional black and white pattern.\footnote{One problem is whether the supergravity solution is determined uniquely by the boundary black and white pattern or not. We expect that this is true because of the analogy of the case of \cite{Lin:2004nb}. However we could not show it in this paper.}

For example, $AdS_5\times S^5$ is a solution that satisfy the ansatz we are considering. The metric of the $AdS_5\times S^5$ can be written as
\begin{align}
 &ds^2=e^{2A}d\check{\Omega}_2^2+R^2(du^2+d\theta^2)+e^{2B}d \hat{\Omega}_2^2
  +e^{2C}d\Omega_4^2, \nn\\
 &e^{A}=R\cosh u,\qquad e^{B}=R\sinh u,\qquad e^{C}=R\sin \theta,\nn\\
 &u \ge 0,\qquad 0\le\theta\le\pi,\qquad R=(4\pi g_s N)^{1/4},
\end{align}
where $d\check{\Omega}_2^2,d \hat{\Omega}_2^2,d\Omega_4^2$ are the metric of unit $AdS_2$, $S^2$, $S^4$ respectively. In this metric, $y:=e^{B}e^{C}=R^{2}\sinh u \sin \theta$. $x$ is defined by $dx=\dt y$, so $x$ is written as $x=-R^2 \cosh u \cos \theta$ in the $AdS_5\times S^5$ solution. On the $x$-axis, $S^2$ shrinks in the region $-R^2\le x \le R^2$, and $S^4$ shrinks in the region $x\le -R^2$ or $R^2 \ge x$. One can draw a diagram like figure \ref{fig-ads} for $AdS_5\times S^5$ geometry.

\subsection{Relation to the Gaussian matrix model}

It is conjectured in \cite{Erickson:2000af,Drukker:2000rr} that the vacuum expectation value of a circular Wilson loop is calculated by the Gaussian matrix integral. For example, the following partition function corresponds to the Wilson loop of trivial representation i.e. identity operator.
\begin{align}
 Z=\int dM \exp\left(
-\frac{1}{\hbar}\tr[M^2]
\right).\label{partition-function}
\end{align}
Here $M$ is a $N\times N$ Hermitian matrix and the measure of integral $\int dM$ is a component-wise one.

We claim that the $x$-axis in our geometry corresponds to the eigenvalue space of the matrix model. In order to see this correspondence, let us first give a short review the ``steepest decent'' method to evaluate the integral \eqref{partition-function}. 

By diagonalization of the matrix $M$, the integral can be rewritten as a integral of the eigenvalues $\lambda_i,\ (i=1,\dots,N)$ as
\begin{align}
 Z=\int (\prod_{i=1}^{N}d\lambda_i) 
  \prod_{1\le i < j\le N}(\lambda_i-\lambda_j)^2
\exp\left(
-\frac{1}{\hbar}\sum_{i=1}^{N}\lambda_i^2
\right), \label{integral-eigenvalue}
\end{align}
where $\prod_{1\le i < j\le N}(\lambda_i-\lambda_j)^2$ is the Jacobian. This is the square of the Vandermonde determinant. We evaluate \eqref{integral-eigenvalue} by the saddle point. In that case, due to the Vandermonde determinant, there is repulsive force between the eigenvalues. The task is to calculate the classical distribution of the eigenvalues.

In order to obtain the distribution, it is convenient to consider the resolvent $\omega(z):=\tr\left[\frac{1}{M-z}\right]=\sum_{i}\frac{1}{\lambda_i-z}$. Due to the equation of motion for $\lambda_j$'s, $\omega(z)$ at the saddle point satisfies the following differential equation
\begin{align}
 0=\frac{2N}{\hbar}+\frac{2}{\hbar}z\omega(z)+\omega(z)^2-\omega'(z).
\label{resolvant}
\end{align}
In the large $N$ limit, the last term $\omega'(z)$ is negligible. Hence eq.\eqref{resolvant} becomes an algebraic equation and we can solve it easily.
\begin{align}
 \omega(z)=-\frac{z}{\hbar}\pm\sqrt{\frac{z^2}{\hbar^2}-\frac{2N}{\hbar}}.
\label{exp-res}
\end{align}
This function has a cut between $z=\pm\sqrt{2\hbar N}$. The density of the eigenvalues is expressed as
\begin{align}
 \rho(\lambda)&=\frac{1}{2\pi i}( \omega(\lambda-i\varepsilon)-\omega(\lambda+i\varepsilon))\nn\\
&=
\begin{cases}
 0, & (\lambda<-\sqrt{2\hbar N}\ \text{or}\ \lambda>\sqrt{2\hbar N}),\\
\frac{1}{\pi}\sqrt{\frac{2N}{\hbar}-\frac{\lambda^2}{\hbar^2}},
& (-\sqrt{2\hbar N} \le \lambda\le\sqrt{2\hbar N}).
\end{cases}
\end{align}

We can compare the $x$-axis in the $AdS_5\times S^5$ geometry and the eigenvalue of the Gaussian matrix model. In the figure \ref{fig-ads} , the length of the black segment is $2\sqrt{4\pi g_s N}$. In the matrix model partition function, the length of the cut, i.e. the distance between the smallest eigenvalue and the largest eigenvalue is $2\sqrt{2\hbar N}$. {\em These two are completely the same if we identify $\hbar=2\pi g_s$.}

\begin{figure}
 \begin{center}
\begin{tabular}{cc}
  \includegraphics[width=200pt]{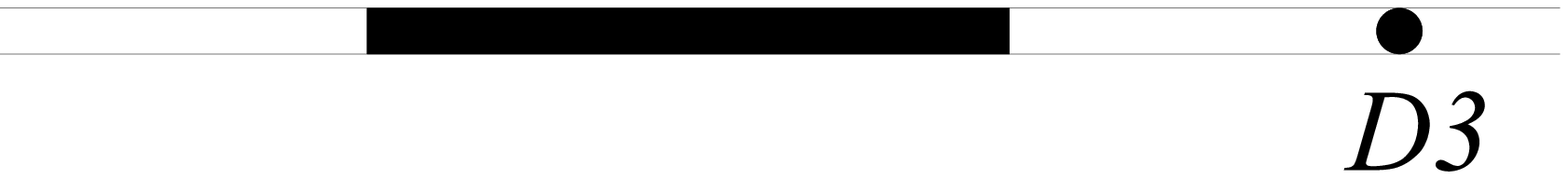}&
  \includegraphics[width=200pt]{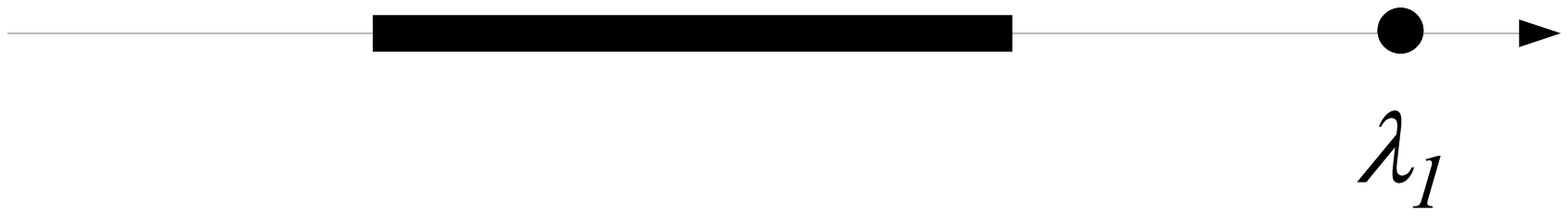}
  \\
(A) & (B)
\end{tabular}
 \end{center}
\caption{Figure (A) expresses the $x$-axis of the $AdS_5\times S^5$. The $AdS_2\times S^2$ D3-brane looks as a point like object on the $x$-axis. It sits on the point $x=\sqrt{4\pi g_s N+\frac{k^2(2\pi g_s)^2}{4}}$. On the other hand, figure (B) represents the eigenvalue distribution of the matrix model. The black bar is the $\lambda_2,\dots,\lambda_N$. Only $\lambda_1$ separates from the other eigenvalues. Its position is $\lambda_1=\sqrt{2\hbar N+\frac{k^2\hbar^2}{4}}$.}\label{particle}
\end{figure}

Next, let us turn to another check. We will consider the operator $\frac{1}{N}\tr[U^k],\quad U:=P\exp(i\int dx^0(A_0+\vp_4))$, for the positive integer $k$. According to \cite{Drukker:2005kx}, the operator $\frac{1}{N}\tr[U^k]$ corresponds to an $AdS_2\times S^2$ D3-brane with $k$ unit of electric flux. This $AdS_2\times S^2$ D3-brane is an analogue of a giant graviton\cite{McGreevy:2000cw,Grisaru:2000zn,Hashimoto:2000zp}. In our base 2-dimensional space, this D3-brane sit at a point on the $x$-axis. The position of this $AdS_2\times S^2$ D3-brane is
\begin{align}
 x=\sqrt{4\pi g_s N+\frac{k^2(2\pi g_s)^2}{4}},\label{ads-particle}
\end{align}
as shown in figure \ref{particle} (A).

On the other hand in the matrix model side, the vacuum expectation value of $\frac{1}{N}\tr[U^k]$ can be calculated by the matrix model as
\begin{align}
 \left\langle\frac{1}{N}\tr[U^k]\right\rangle=\frac{1}{Z}\int dM 
\frac{1}{N}\tr[e^{kM}]
\exp\left(
-\frac{1}{\hbar}\tr[M^2]
\right).
\end{align}

Diagonalizing the matrix $M$ leads to
\begin{align}
 \frac{1}{Z}\int (\prod_{i=1}^{N}d\lambda_i) 
  \prod_{1\le i < j\le N}(\lambda_i-\lambda_j)^2
\sum_{i=1}^{N}\frac{1}{N}e^{k\lambda_i}
\exp\left(
-\frac{1}{\hbar}\sum_{i=1}^{N}\lambda_i^2
\right)\nn\\
=\frac{1}{Z}\int (\prod_{i=1}^{N}d\lambda_i) 
  \prod_{1\le i < j\le N}(\lambda_i-\lambda_j)^2
e^{k\lambda_1}
\exp\left(
-\frac{1}{\hbar}\sum_{i=1}^{N}\lambda_i^2
\right).
\end{align}
Here the equations of motion for $\lambda_2,\dots,\lambda_{N}$ are the same as before. Since $N$ is large now, we use the solution for $\lambda_2,\dots,\lambda_{N}$ neglecting $\lambda_1$ and take it as a background for $\lambda_1$. In this case, the equation of motion for $\lambda_1$ becomes
\begin{align}
 0=\frac{2}{\hbar}\lambda_1-2\sum_{i=2}^{N}\frac{1}{\lambda_1-\lambda_i}-k.
\label{lambda1}
\end{align}
The second term of the right-hand side of the above equation becomes $-2\omega(\lambda_1)$ of $(N-1)\times (N-1)$ matrix model. We neglect the difference of $N$ and $N-1$ and replace this term with the expression of \eqref{exp-res}. Then, we can solve \eqref{lambda1} as
\begin{align}
 \lambda_1=\sqrt{2\hbar N+\frac{k^2\hbar^2}{4}},\label{mm-particle}
\end{align}
as shown in figure \ref{particle} (B). {\em The positions of the particles in \eqref{ads-particle} and \eqref{mm-particle} completely match.}
\subsection{The Maya diagram and the Young diagram}
In this subsection, we propose a rule of correspondence between the geometry and the Young diagram: the label of the Wilson line\footnote{Kazuo Hosomichi kindly explained to me about the basic idea of this subsection. I appreciate it very much.}.

First, let us explain how to discretize the pattern of $x$-axis. Since the black segment expresses a $S^5$, we can replace a black segment with a sequence of black dots of the same number as the 5-form flux through the $S^5$. As the same way we will replace white segment with white dots of the same number as the 3-form flux through the $S^3$. The semi-infinite white lines should be replaced by infinite number of white dots. Then we will obtain a kind of maya diagram.

This is not the ordinary maya diagram. In ordinary maya diagram, every dot is black after a certain position in the left, while every dot is white after a certain position in the right. This kind of maya diagram corresponds to a general Young diagram. On the other hand, in our maya diagram here, every dot is white after a certain position in the both direction. This kind of maya diagram is associated to a Young diagram with rows less than $N$: the number of black dots.

\begin{figure}
 \begin{center}
  \includegraphics[width=300pt]{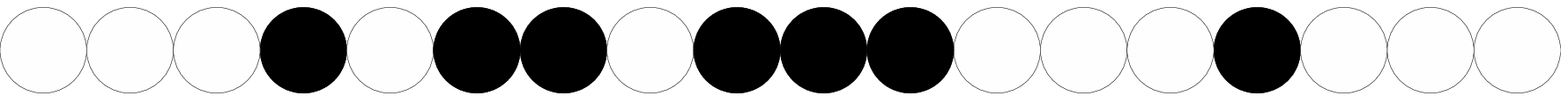}
 \end{center}
\caption{An example of maya diagram.}
\label{mayaex}
\end{figure}
\begin{figure}
 \begin{center}
\begin{tabular}{cc}
  \includegraphics[height=120pt]{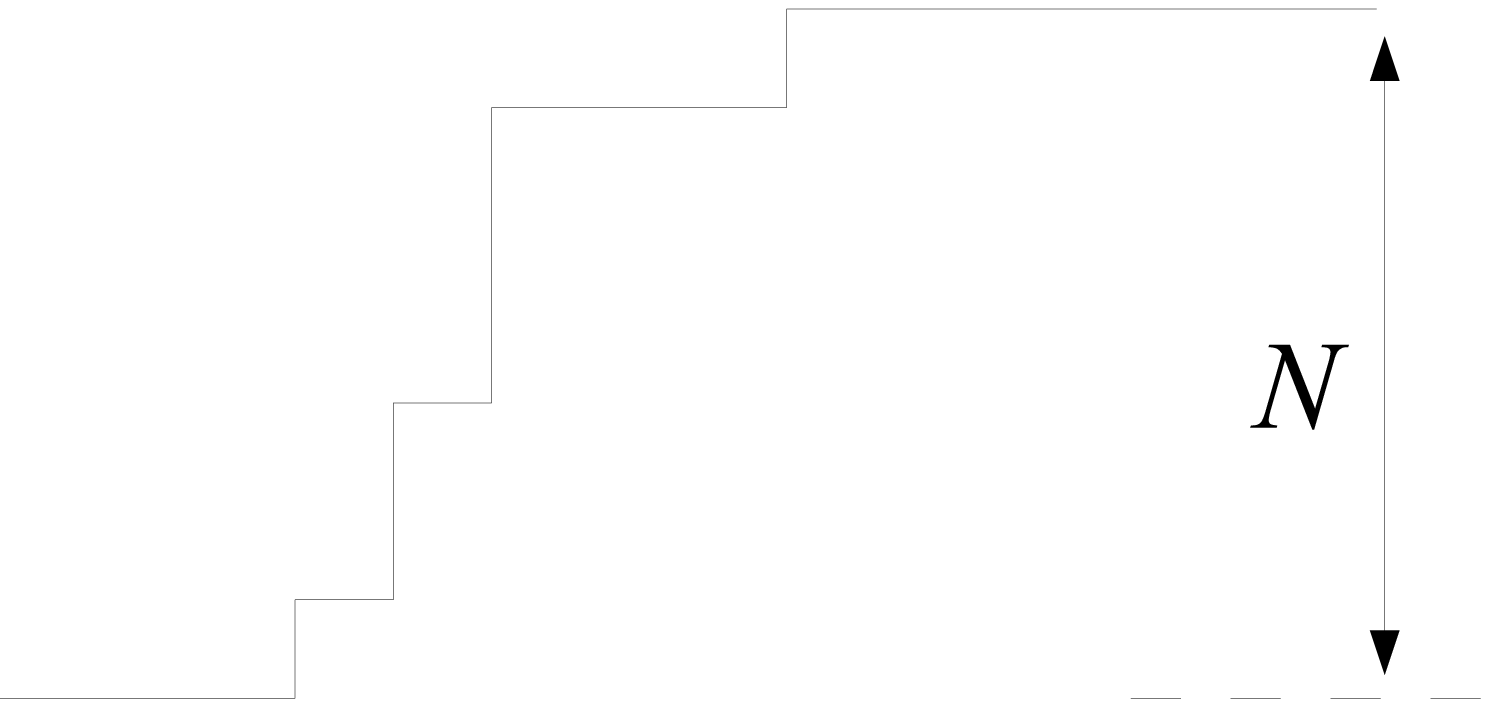}&
  \includegraphics[height=120pt]{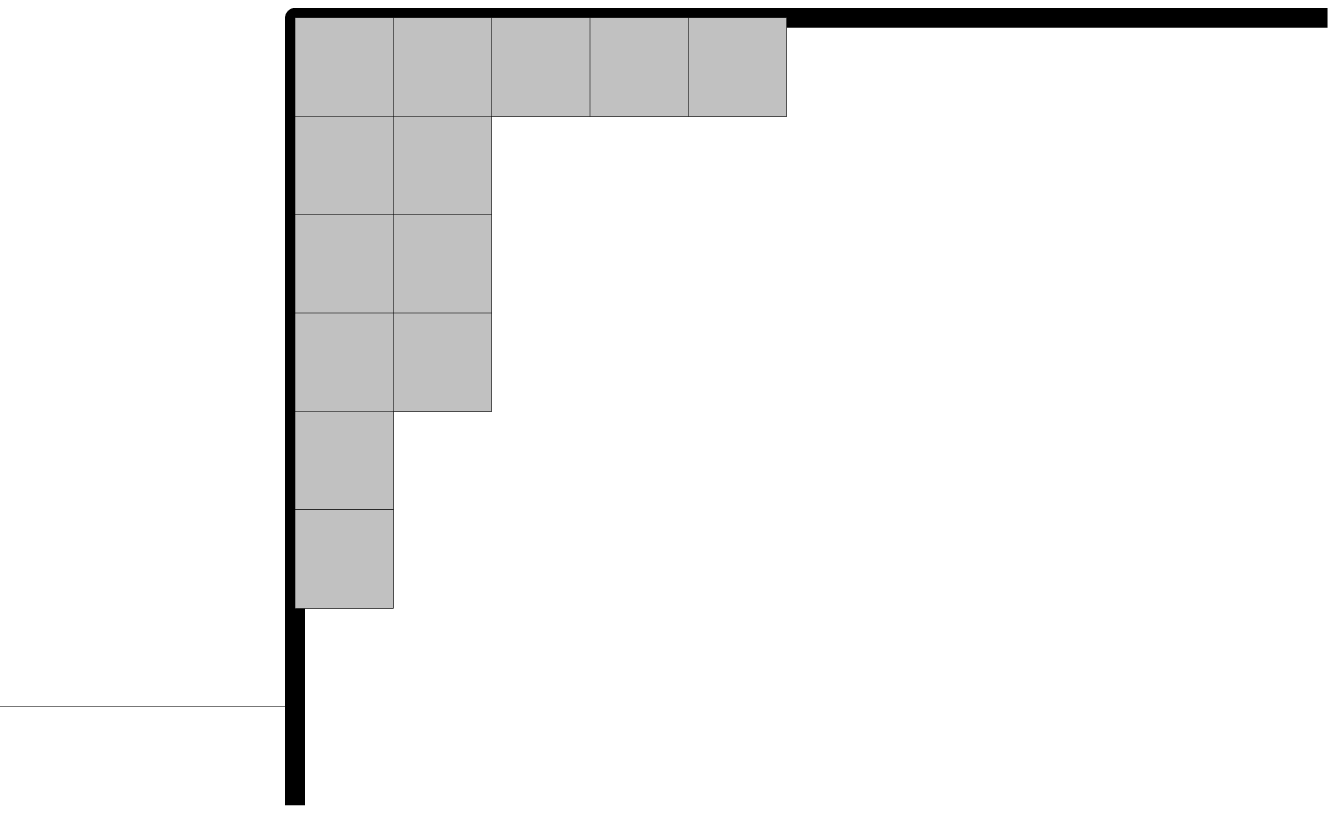}
  \\
(A)&(B)
\end{tabular}
 \end{center}
\caption{Correspondence between maya diagrams and Young diagrams. Here we show the Young diagram which corresponds to the example of figure \ref{mayaex}. The figure of (A) is the line made by replacing a white dot with horizontal segment and a black dot with vertical segment of unit length. The ``height'' of this figure is equal to the number of black dots $N$. The figure (B) shows how to make the Young diagram from the figure (A).}
\label{figmaya}
\end{figure}
We associate the maya diagram to Young diagram as the following way. First, as the same way as the ordinary maya diagram, replace a white dot with ``go right'' and a black dot with ``go up'', and draw a path as figure \ref{figmaya} (A). Next, cut the upper area of the path by the line like figure \ref{figmaya} (B). Then we obtain the Young diagram with rows less then $N$.

If we take the matrix model eigenvalue picture into account, we find that this Young diagram does not correspond to the label of ``representation basis'' of the Wilson line operators. Actually this Young diagram corresponds to the label of the following ``monomial basis'' of the Wilson line operators.

In order to explain this basis, let us define the $N\times N$ unitary matrix $U:=P\exp(i\int dx^0(A_0+\vp_4))$, whose eigenvalues are $u_1,\dots,u_N$. It is convenient to express a Young diagram with rows less then $N$ by $(\mu_1,\dots, \mu_N )$, where $\mu_j$ is the number of boxes in the $j$-th row. The monomial basis for the Young diagram $\mu$ is $m_{\mu}=u_1^{\mu_1}u_2^{\mu_2} \dots u_N^{\mu_N}+(\text{sym})$, where (sym) means the terms for symmetrization. On the other hand, the representation basis is expressed by the Schur polynomial $\Tr_{R}[U]=S_{R}(u_1,\dots,u_N)$ for a Young diagram $R$. Of course we can express one basis in terms of some linear combination of the other.

Let us take the Young diagram $(k,0,0,\dots)$ for an example. The symmetric monomial becomes
\begin{align}
 m_{(k,0,0,\dots)}=u_1^{k}+(sym)=\tr[U^k].
\end{align} 
The corresponding Maya diagram looks like figure \ref{particle} (B). This operator corresponds to the $AdS_2\times S^2$ D3-brane in the AdS side.

Let us comment on the translation and the left-right flip of the maya diagram. In the geometry side, the system is symmetric under the translation and the left-right flip of the maya diagram. How can we see these symmetries in the label of Wilson line operator? First, translation means inserting or removing the determinant $\det U=u_1\dots u_N$. In the language of the Young diagram, translation means $(\mu_1,\dots,\mu_N)\to (\mu_1+a,\dots,\mu_N+a)$ for an integer $a$. Usually using this degrees of freedom we can set $\mu_{N}=0$ and draw a Young diagram with rows less than $N$. As for the left-right flip, it corresponds to the complex conjugation. The monomial basis satisfy $m_{\mu}(u^{*})=m_{-\mu}(u)$, since each $u_j$ is a phase, that is to say, $u_j^{*}=u_j^{-1}$. Here Young diagram $(-\mu)$ means $(-\mu)_j=-\mu_{N-j}$. Also in this case, using translation symmetry, we can draw a Young diagram with rows less than $N$.

\section{1/2 BPS Geometry in M-theory}
\label{M}
In this section, we discuss the similar problem in M-theory. For example, in the 6-dimensional (2,0) superconformal field theory, we have a kind of surface operators. It preserves half of the supersymmetry if the shape is flat. We first consider the symmetry preserved by this kind of operators, and make the ansatz. Then, we study the supersymmetry conditions and derive the necessary conditions. We also show a few examples.

\subsection{Ansatz for the surface operators in 6-dimensional SCFT and 
  3-dimensional SCFT}

In the AdS/CFT correspondence context, 6-dimensional (2,0) SCFT corresponds to M-theory on the $AdS_7\times S^4$ background. The surface operator corresponds to a membrane in the probe picture. In this picture, we can easily see that the bosonic symmetry which is preserved by this surface operator (or the membrane) is $SO(2,2)\times SO(4)\times SO(4)$.

We can also consider a 3-dimensional SCFT which corresponds to $AdS_4\times S^7$. In this theory, we can also consider the wall like defect operators, which corresponds to $AdS_3\times S^3$ M5-brane. The symmetry which is preserved by this operator is also $SO(2,2)\times SO(4)\times SO(4)$, which can be seen by the probe picture.

Let us turn to making the ansatz that respects the symmetry $SO(2,2)\times SO(4)\times SO(4)$. The metric can be written in the form 
\begin{align}
 ds^2=e^{2A}d\check{\Omega}_3^2+ds_2^2+e^{2B}d \hat{\Omega}_3^2
  +e^{2C}d\Omega_3^2, \label{metric-ansatz-m}
\end{align}
where $d\check{\Omega}_3^2,d \hat{\Omega}_3^2,d\Omega_3^2$ are the metrics of unit $AdS_3$, $S^3$, $S^3$ respectively. $ds_2^2$ in \eqref{metric-ansatz-m} is a general 2-dimensional metric to be determined. As in the IIB case, we call this 2-dimensional space ``base space.'' $A,B,$ and $C$ in \eqref{metric-ansatz-m} are functions on the base space.

As for the flux, the most general ansatz which preserves the $SO(2,2)\times SO(4)\times SO(4)$ symmetry is
\begin{align}
 G_4=6 F E^{0}E^{1}E^{2}+6J E^{5}E^{6}E^{7}+6K E^{8}E^{9}E^{10},
\end{align}
where $E^{a},a=0,\dots,10$ are the vielbein and $F,J,K$ are 1-forms on the base 2-dimensional space. $(E^{0},E^{1},E^{2})$ are the vielbein of $AdS_3$ part, $(E^{5},E^{6},E^{7})$ are the vielbein of one $S^{3}$ part, and $(E^{8},E^{9},E^{10})$ are the vielbein of the other $S^{3}$ part.

We study the supersymmetry conditions under this ansatz in the next subsection. There are some related works. In \cite{Martelli:2003ki}, they have obtained some necessary and sufficient conditions for $AdS_3\times X$ geometry. So our problem is the special case of them. In \cite{Bena:2004jw}, they obtained the solutions which includes $\Rb^{1,2}\times S^3 \times S^3$. Our problem is to obtain the geometry that includes $AdS_3$ instead of $\Rb^{1,2}$.
\subsection{The gravitino condition and the spinor bilinears}
In this subsection, we use the following convention of 11-dimensional gamma matrices.
\begin{align}
 &\Gamma^{0}=\gamma_5\otimes\sigmac^{0}\otimes 1\otimes 1\otimes \sigma_1, \qquad
\Gamma^{1}=\gamma_5\otimes\sigmac^{1}\otimes 1\otimes 1\otimes \sigma_1, \qquad
\Gamma^{2}=\gamma_5\otimes\sigmac^{2}\otimes 1\otimes 1\otimes \sigma_1, \nn\\
 &\Gamma^{3}=\gamma^{3}\otimes1\otimes 1\otimes 1\otimes 1, \qquad
\Gamma^{4}=\gamma^{4}\otimes1\otimes 1\otimes 1\otimes 1, \nn\\
 &\Gamma^{5}=\gamma_5\otimes1\otimes \sigmah^5\otimes 1\otimes \sigma_2, \qquad
\Gamma^{6}=\gamma_5\otimes1\otimes \sigmah^6\otimes 1\otimes \sigma_2, \qquad
\Gamma^{7}=\gamma_5\otimes1\otimes \sigmah^7\otimes 1\otimes \sigma_2, \nn\\
&\Gamma^{8}=\gamma_5\otimes1\otimes 1\otimes \sigmat^{8}\otimes \sigma_3, \qquad
\Gamma^{9}=\gamma_5\otimes1\otimes 1\otimes \sigmat^{9}\otimes \sigma_3, \qquad
\Gamma^{10}=\gamma_5\otimes1\otimes 1\otimes \sigmat^{10}\otimes \sigma_3.
\end{align}
Here $(\sigma_1,\sigma_2,\sigma_3),(\gamma^{3},\gamma^{4},\gamma_{5}),\ (\sigmah^{5},\sigmah^{6},\sigmah^{7})$, and $(\sigmat^{8},\sigmat^{9},\sigmat^{10})$ are sets of the Pauli matrices. $(\sigmac^{1},\sigmac^{2},\sigmac^{3})$ is also a set of the Pauli matrices and we define $\sigmac^{0}:=i\sigmac^{3}$.

We also use Killing spinors in AdS spacetimes and spheres. For $AdS_3$, they can be written as
\begin{align}
 &\nablac_p \chic^{I}_{a}=\frac i2 a \sigmac_{p}\chic^{I}_{a},\qquad 
 (p=0,1,2,\qquad a=\pm 1,\qquad I=1,2),
\end{align}
where $\nablac$ is the covariant derivative of Levi-Civit\'a connection of unit $AdS_3$. As the same way, there are Killing spinors in $S^3$.
\begin{align}
  &\nablac_p \chih^{J}_{b}=\frac 12 b \sigmah_{p}\chih^{J}_{b},\qquad 
 (p=5,6,7,\qquad b=\pm 1,\qquad J=1,2),
\end{align}
where $\nablac$ is the covariant derivative of Levi-Civit\'a connection of unit $S^3$. We also prepare another set of $S^3$ Killing spinors $\chi^{K}_{c}$ for the other $S^3$.

We can expand the 11-dimensional spinor $\xi$ as $\xi=\sum_{abcIJK} \epsilon_{abcIJK}\otimes \chic^{I}_{a}\otimes \chih^{J}_{b}\otimes\chi^{K}_{c}$. In this expansion, each of the coefficient $\epsilon_{abcIJK}$ is an element of (2 dim spinor)$\otimes \Cb^2$. $\gamma_{m},\ (m=3,4,5)$ act on the (2 dim spinor) part, and $\sigma_{j},\ (j=1,2,3)$ act on the $\Cb^2$ part.

With these notations, we can rewrite the supersymmetry conditions as
\begin{align}
 &0=\left[
ae^{-A}\sigma_1-\delsla A \gamma_5 + 2 \Fs\sigma_1-\Js i\sigma_2-\Ks i\sigma_3
\right]\epsilon,\label{ma}\\
 &0=\left[
ibe^{-B}\sigma_2-\delsla B \gamma_5 -  \Fs\sigma_1+2\Js i\sigma_2-\Ks i\sigma_3
\right]\epsilon,\label{mb}\\
 &0=\left[
ice^{-C}\sigma_3-\delsla C \gamma_5 -  \Fs\sigma_1-\Js i\sigma_2+2\Ks i\sigma_3
\right]\epsilon,\label{mc}\\
 &0=\nabla_{m}\epsilon-\frac{i}{2}\left[
  \Ft_m\sigma_1+\Jt_m i\sigma_2+\Kt_m i\sigma_3\right]\epsilon
 -\gamma_5\left[
  F_m\sigma_1+J_m i\sigma_2+K_m i\sigma_3\right]\epsilon,\label{mm}
\end{align}
where tilde is the 2-dimensional Hodge dual. In these equations, we omit the
index of $\epsilon_{abcIJK}$.

Let us consider the spinor bilinears
\begin{align}
 f_0:=\epsilon^{\dag}\epsilon,\quad f_j:=\epsilon^{\dag}\sigma_j\epsilon,\quad
g_0:=\epsilon^{\dag}\gamma_5\epsilon,\quad g_j:=\epsilon^{\dag}\gamma_5\sigma_j\epsilon.
\end{align}
The derivative of these functions can be calculated by using \eqref{mm}.
\begin{align}
 df_0&=2g_1 F-f_2\Jt-f_3 \Kt,\label{df0m}\\
 df_2&=f_3\Ft-f_0\Jt-2g_1 K,\\
 df_3&=-f_2\Ft+2g_1J-f_0\Kt,\\
 dg_1&=2f_0F-2f_3J+2f_2K.\label{dg1m}
\end{align}
On the other hand, from eqs.\eqref{ma}-\eqref{mc}, we can derive the relations
\begin{align}
 f_0dA&=2g_1 F-f_2\Jt-f_3 \Kt,\label{f0da}\\
 f_3dB&=-f_2\Ft+2g_1J-f_0\Kt,\\
 f_2dC&=f_3\Ft-f_0\Jt-2g_1 K.\label{f2dc}
\end{align}
From \eqref{df0m}-\eqref{dg1m} and \eqref{f0da}-\eqref{f2dc} one finds that $e^{A},e^{B},e^{C}$ are proportional to $f_0,f_3,f_2$ respectively. By adjusting the normalization of $\epsilon$, we can set $f_0=e^{A}$, while $f_2$ and $f_3$ can be written with constant coefficients $p,q$ as
\begin{align}
 f_3=pe^{B},\qquad f_2=qe^{C}.
\end{align}

Next, let us show $y:=e^{A+B+C}$ is a harmonic function on the base 2-dimensional space. Summing up eqs.\eqref{ma}-\eqref{mc} gives
\begin{align}
 0=\left[
 ae^{-A}\sigma_1+ibe^{-B}\sigma_2+ice^{-C}\sigma_3
 -\delsla (A+B+C)\gamma_5
\right]\epsilon.\label{mabc}
\end{align}
From this equation and its hermitian conjugation, we obtain
\begin{align}
 \dt y=-be^{C}P^{(2)}-ce^{B}P^{(3)},\label{dty}
\end{align}
where $P^{(j)}$'s are spinor bilinears defined as
\begin{align}
 P^{(0)}_{m}:=\epsilon^{\dag}\gamma_{m}\epsilon,\qquad
 P^{(j)}_{m}:=\epsilon^{\dag}\sigma_{j}\gamma_{m}\epsilon,
\end{align}
and 1-form is defined for example $P=P_m E^{m}$. In order to show the right-hand side of \eqref{dty} is a closed 1-form, let us first show that $e^{B}P^{(3)}$ is closed. From eq.\eqref{mb}, we have the relation
\begin{align}
 &0=be^{-B}P^{(3)}+\frac12 dg_1+g_1dB+3f_3J.
\end{align}
This equation and eq.\eqref{dg1m} read 
\begin{align}
 0=be^{B}P^{(3)}+d\left(\frac12e^{2B}g_1\right)+3pe^{3B}J.
\end{align}
The second term of the left-hand side is exact 1-form, and the last term is closed because of the Bianchi identity for the 4-form field strength of the 11-dimensional supergravity. So we can conclude that $e^{B}P^{(3)}$ is a closed 1-form. As the same way, we can show that $e^{C}P^{(2)}$ is closed. Thus $\dt y$ is closed because of eq.\eqref{dty}. We can define the other coordinate $x$ by $dx=\dt y$. 

We can express the norm $|dy|^2$ by $A,B,C$. Multiply \\
 $\left[
 ae^{-A}\sigma_1+ibe^{-B}\sigma_2+ice^{-C}\sigma_3
 +\delsla (A+B+C)\gamma_5 \right]$
 to \eqref{mabc}, and we obtain
\begin{align}
 0=e^{-2A}-e^{-2B}-e^{-2C}-|dy|^2/y^2.
\end{align}
This equation reads $|dy|^2=-e^{2B+2C}+e^{2A+2B}+e^{2A+2C}$. Hence the metric of 2-dimensional space can be written as
\begin{align}
 ds_2^2=\frac{1}{-e^{2B+2C}+e^{2A+2B}+e^{2A+2C}}(dy^2+dx^2).
\end{align}

From eq.\eqref{mabc}, we can show that $P^{(1)}=0$. This fact and the Fierz identity read
\begin{align}
 0=-f_0^2+f_2^2+f_3^2+g_1^2.
\end{align}
Eq.\eqref{mabc} also lead to the relation between constants,
\begin{align}
 0=a-bp+cq.
\end{align}

In summary, we obtain the necessary conditions of supersymmetry.
\begin{align}
 &ds_2^2=\frac{1}{-e^{2B+2C}+e^{2A+2B}+e^{2A+2C}}(dy^2+dx^2),\\
 &6F=4\frac{df_0}{g_1}-\frac{f_0dg_1}{g_1^2}+\frac{2}{g_1^2}(f_2\dt f_3
  -f_3 \dt f_2),\\
 &6J=4\frac{df_3}{g_1}-\frac{f_3dg_1}{g_1^2}+\frac{2}{g_1^2}(-f_0\dt f_2+
  f_2\dt f_0),\\
 &6K=-4\frac{df_2}{g_1}+\frac{f_2dg_1}{g_1^2}+\frac{2}{g_1^2}(-f_0\dt f_3
 +f_3\dt f_0),
\end{align}
where 
\begin{align}
 f_0=e^{A},\qquad f_3=pe^{B},\qquad f_2=qe^{C},\qquad (p,q:\text{constants}),\qquad
 g_1=\sqrt{f_0^2-f_2^2-f_3^2}.
\end{align}

The functions $A,B,C$ need to satisfy some differential equations. For example, they should satisfy at least the following equations derived from the Bianchi identity.
\begin{align}
 d\left[\frac{y}{g_1^2}e^{2A}\dt(B-C)\right]=0,\qquad
 d\left[\frac{y}{g_1^2}e^{2B}\dt(C-A)\right]=0,\qquad
 d\left[\frac{y}{g_1^2}e^{2C}\dt(A-B)\right]=0.
\end{align}

To obtain all the conditions and show it to be sufficient is a future problem.
\subsection{Continuous maya diagram and examples}
As in the IIB case, the base 2-dimensional space has a boundary defined by $y=0$. On this boundary $y=e^{A+B+C}=0$ is satisfied. In order to make the geometry regular, one of $e^B$ and $e^C$ vanishes at a point on the boundary but the other should not vanish at that point.  Let us make a 1-dimensional pattern as the same way as the IIB case; Paint the point $e^{B}=0$ in black, and paint the point $e^{C}=0$ in white.

Here, we mention some examples. $AdS_7\times S^4$, $AdS_4\times S^7$ and $AdS_3
\times S^3\times R^4\times S^1$.

First let us see the $AdS_7\times S^4$ solution. It can be written as
\begin{align}
 &ds_{2}^{2}=R^{2}(du^2+d\theta^2),\nn\\
 &e^{A}=R \cosh u,\qquad e^{B}=R\sinh u,\qquad e^{C}=\frac{R}{2}\sin 2\theta,\\
 &y:=e^{A+B+C}=\frac{R^3}4 \sinh 2u \sin 2\theta,\qquad
  x=-\frac{R^3}4 \cosh 2u \cos 2\theta,
\end{align}
where $R$ is a constant. The $x$-axis of this solution looks just the same as
figure \ref{fig-ads}.

\begin{figure}
 \begin{center}
  \includegraphics[width=300pt]{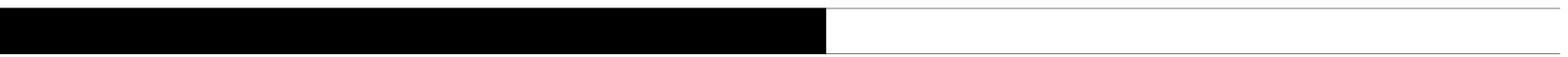}
 \end{center}
\caption{$x$-axis of $AdS_4\times S^7$. It is painted in black $x<0$ and in 
 white $x>0$.}
\label{fig-half}
\end{figure}

Second let us see the $AdS_4 \times S^7$ solution
\begin{align}
 &ds_{2}^{2}=R^{2}(du^2+d\theta^2),\nn\\
 &e^{A}=\frac{R}{2} \cosh 2u,\qquad e^{B}=R\cos \theta,\qquad e^{C}=R\sin\theta,\\
 &y=\frac{R^3}{4}\cosh 2u \sin 2\theta,\qquad
 x=-\frac{R^3}{4}\sinh 2u \cos 2\theta.
\end{align}
The pattern of $x$-axis of this solution looks like figure \ref{fig-half}.

Finally let us see the $AdS_3\times S^3\times R^4\times R^1$
\begin{align}
&ds_2^2=du^2+dv^2,\\
 &e^{A}=R,\qquad e^{B}=R,\qquad e^{C}=u,\\
 &y=R^2 u, \qquad x=R^2 v.
\end{align}
In this solution, the pattern of $x$-axis is just a white line.
\section{Conclusions and discussions}
\label{conclusion}
In this paper, we investigate the geometry that corresponds to the half BPS straight Wilson line. In this geometry, we can see the distribution of eigenvalues of the Gaussian matrix model, which is supposed to describe the theory of half BPS Wilson lines. We also consider similar half BPS geometry in M-theory.

\begin{figure}
 \begin{center}
  \includegraphics[width=300pt]{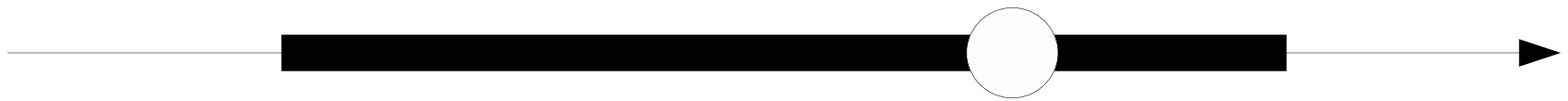}
 \end{center}
\caption{The eigenvalue distribution with a hole. The thick line represents where the eigenvalues distributed. The white dot represents the hole where the eigenvalue is not distributed. In the $AdS_5\times S^5$ picture, the thick line is the place on the $x$-axis where the $S^4$ has finite size. So we can guess a hole corresponds to $AdS_2\times S^4$ D5-brane. On the other hand in Yang-Mills theory, the hole corresponds to the Wilson line of anti-symmetric representation.}
\label{fig-hole}
\end{figure}

In this paper, we identify an isolated eigenvalue in matrix model to an $AdS_2\times S^2$ D3-brane. Then, what is the isolated ``hole'' in the matrix model? (See figure \ref{fig-hole}) We guess that an $AdS_2 \times S^4$ D5-brane corresponds to a hole in the eigenvalue distribution of the Gaussian matrix model. To investigate this probe brane and its relation to the Wilson line operator of the anti-symmetric representation is a future problem.

We could not obtain so far necessary and sufficient condition of supersymmetric background satisfying the ansatz studied here. It is an important problem to obtain the necessary and sufficient condition and make nontrivial examples of supergravity solutions. 

The solutions of the other kind of extended defect operators also seem interesting to study. For example, the general Wilson-'t Hooft lines will corresponds to the most general ansatz that preserves SL(2,R)$\times$SO(3)$\times$SO(5). In probe picture, they correspond to $(p,q)$-strings or $AdS_2\times S^2$ D3-branes with electro-magnetic flux or $AdS_2\times S^4$ $(p,q)$ 5-brane with electric flux. Another interesting class of operators is wall-like defect operators. The probe picture is typically $AdS_4\times S^2$ D5-brane \cite{Karch:2000ct,Karch:2000gx}. This class of operators preserve $SO(2,3)\times SO(3)\times SO(3)$ and half of the supersymmetry.

%%%%%%%%%%%%%%%%%%%%%%%%%%%% acknowledgment %%%%%%%%%%%%%%%%%%%%%%%%%%%%%%
\subsection*{Acknowledgment}
I would like to thank Mohab Abou-Zeid, Mitsuhiro Arikawa, Tohru Eguchi, Kazuo Hosomichi, Yosuke Imamura, So Matsuura, Sanefumi Moriyama, Nikita Nekrasov, Kazutoshi Ohta, Jeong-Hyuck Park, Gordon W. Semenoff, Tadashi Takayanagi, Jan Troost,  Tatsuya Tokunaga,  Marcel Vonk, and Shing-Tung Yau for useful discussions and comments, and Juan Maldacena for e-mail correspondence. I am also grateful for the hospitality of Theoretical Physics Group of RIKEN, Yukawa Institute of Theoretical Physics, and Theoretical Group of KEK. Discussions during the YITP workshop YITP-W-05-2 on ``Fundamental Problems and Applications of Quantum Field Theory'' were useful to complete this work. This work was supported in part by the European Research Training Network contract 005104 ``ForcesUniverse.''

\appendix
\section{Convention for supergravity}
\subsection{Gamma matrices}
We use 10- and 11-dimensional gamma matrices satisfying
\begin{align}
& \{\Gamma^M,\Gamma^N\}=2\eta^{MN},\qquad \eta^{MN}=\text{diag}(-1,+1,\dots,+1)
,\qquad M,N=0,\dots,10,\\
&\Gamma^{0}\Gamma^{1}\dots \Gamma^{9}\Gamma^{10}=+1.
\end{align}
Hodge dual for a $p$-form $G$ is
defined by the anti-symmetric tensor $\varepsilon^{M_0\dots M_n}$
, $n=9$ or $n=10$ satisfying $\varepsilon^{01\dots n}=+1$ as
\begin{align}
 (*G)_{N_1N_2\dots N_{n+1-p}}:=\frac{1}{p!}\varepsilon_{N_1N_2\dots N_{n+1-p} M_1 M_2\dots M_p} G^{M_1\dots M_p}.
\end{align}
We also use the slash notation defined as
\begin{align}
 \sla{G}:=\frac{1}{p!}G_{M_1\dots M_p}\Gamma^{M_1\dots M_p}.
\end{align}

\subsection{IIB supergravity}
We use the following convention for IIB supergravity. This theory contains the metric $g_{MN}$, the dilaton $\phi$, NSNS 3-form field strength $H_3$, and RR field strength $G_1,G_3,G_5$ as the bosonic fields. This theory also contains gravitino $\psi_{M}$ which is the pair of vectorial Majorana-Weyl spinors of positive chirality $\Gamma_{10} \psi_M=+\psi_M$ as well as dilatino $\lambda$ which is the pair of Majorana-Weyl spinors of negative chirality $\Gamma_{10} \lambda=-\lambda$ as the fermionic fields. The parameter of the SUSY transformation is a doublet of the Majorana-Weyl spinors $\xi=(\xi_1,\xi_2)$ with positive chirality.  $\tau_j,\ (j=1,2,3)$ are a set of Pauli matrices that acts as $(\tau_j \xi)_{\alpha}=(\tau_j)_{\alpha\beta}\xi_{\beta}$. The SUSY transformation for gravitino $\psi_M$ and dilatino $\lambda$ for a bosonic configuration is
\begin{align}
 &\delta \psi_{M}=\nabla_{M}\xi+\frac 18 H_{MAB}\Gamma^{AB}\tau_3 \xi
  +\frac{e^{\phi}}{8}(i\sla{G}_1\tau_2-\sla{G}_3\tau_1
                    +\frac{i}{2}\sla{G}_5\tau_2) \Gamma_{M}\xi,
\label{gravitino-condition}\\
 &\delta \lambda=(\delsla \phi)\xi+\frac12 \sla{H}_3\tau_3\xi
   +e^{\phi}(-i\sla{G}_1\tau_2+\frac12 \sla{G}_3 \tau_1)\xi.
\label{dilatino-condition}
\end{align}

Bianchi identities for the form fields are
\begin{align}
 dH_3=0,\qquad dG_1=0,\qquad dG_3=H_3\wedge G_1,\qquad dG_5=H_3\wedge G_3.
\end{align}
$G_5$ satisfies the self-duality $G_5 = -*G_5$. The equations of motion for form fields are
\begin{align}
 d(e^{-2\phi}*H_3)=-G_3\wedge G_5+G_1\wedge *G_3,\qquad
 d*G_1=*G_3\wedge H_3,\qquad d *G_3 = -G_5\wedge H_3.
\end{align}

\subsection{11-dimensional supergravity}
This theory contains the metric $g_{MN}$ and 4-form field strength $G_4$ as the bosonic fields. This theory also contains gravitino $\psi_{M}$ which is a vectorial Majorana spinors as the fermionic fields. The parameter of SUSY transformation $\xi$ is a Majorana spinor in 11 dimensions. The SUSY transformation of gravitino in a bosonic configuration is 
\begin{align}
 \delta \psi_{M}=\nabla_{M}\xi+\frac{1}{12\times 
 4!}G_{NPQR}\Gamma_{M}{}^{NPQR}\xi-\frac{1}{6\times 3!}G_{MPQR}\Gamma^{PQR}\xi.
\end{align}
The Bianchi identity and the equation of motion for the 4-form are
\begin{align}
 dG_4=0,\qquad d*G_4=G_4\wedge G_4.
\end{align}

%%%%%%%%%%%%%%%%%%%%%%%%%%%%% bibliography %%%%%%%%%%%%%%%%%%%%%%%%%%%%%%%%%%%
%\nocite{*}
\providecommand{\href}[2]{#2}\begingroup\raggedright\endgroup

\end{document}